\documentclass[aps,prb,reprint,twocolumn,floatfix,superscriptaddress,nofootinbib]{revtex4}
\usepackage{graphicx}
\usepackage{amssymb}
\usepackage{amsmath}
\usepackage{lmodern}
\usepackage{color}
\usepackage{hyperref}
\usepackage{empheq}
\usepackage{epstopdf}
\begin{document}

\title
{Doping dependence of low-energy charge collective excitations in high-T$_c$ cuprates}


\author{V.\,M.~Silkin}

\affiliation{Donostia International Physics Center, P. de Manuel Lardizabal 4, 20018 San
Sebasti\'an/Donostia, Basque Country, Spain}
\affiliation{Departamento de Pol\'{\i}meros y Materiales Avanzados: F\'{\i}sica, Qu\'{\i}mica y Tecnolog\'{\i}a, Facultad de Ciencias Qu\'{\i}micas, Universidad del Pa\'{\i}s Vasco UPV/EHU,
Apartado 1072, 20080 San Sebasti\'an/Donostia, Basque Country, Spain}
\affiliation{IKERBASQUE, Basque Foundation for Science, 48009 Bilbao, Basque Country, Spain}

\author{D.\,V.~Efremov}

\affiliation{Leibniz Institute for Solid State and
Materials Research IFW Dresden, Helmholtzstra\ss e 20, 01069 Dresden, Germany}

\author{M.\,Yu.~Kagan}

\affiliation{National Research University Higher School of Economics, Myasnitskaya Ulitsa 20, 101000, Moscow, Russia}
\affiliation{P. L. Kapitza Institute for Physical Problem, Russian Academy of Sciences, Ulitsa Kosygina 2, 119334, Moscow, Russia}



\begin{abstract}

In this study, we analyze the dielectric function of high-Tc cuprates as a function of doping level, taking into account the full energy band dispersion within the CuO$_2$ monolayer.
 In addition to the conventional two-dimensional (2D) gapless plasmon mode, our findings reveal the existence of three anomalous branches within the plasmon spectrum.
  Two of these branches are overdamped modes, designated as hyperplasmons, and the third is an almost one-dimensional plasmon mode (1DP).
We derive an analytic expression for dynamic part of the response function. Furthermore, we investigated the effect of the doping on these modes.
 Our analysis demonstrates that in the doping level range close to the optimal doping level, the properties of all three modes undergo a significant transformation.


\end{abstract}

\maketitle

\section{Introduction}

%
Electronic charge excitations that can occur in a metallic system are well understood within the framework of a free-electron gas (FEG) model. In a three-dimensional (3D) electron system within this model, the low-energy excitations are incoherent electron-hole (e-h) pairs that form a structureless continuum.\cite{li54,pino66}  In addition to the structureless continuum,  charge collective excitations known as plasmons emerge at energies of the order of the conduction electron bandwidth due to the Coulomb interaction.\cite{pibopr52,pibopr53} The resulting plasmon bands are narrow compared to the electronic bands and can be easily detected in optical experiments.
But, due to the high energy of their creation, they are not directly relevant to  most of the phenomena of condensed matter physics occurring at low energy scales, such as electrical conductivity, heat transfer, and many others. Therefore it is widely accepted that only incoherent e-h excitations play any role in these phenomena.
%
%
%

%
In two-dimensionsional (2D) systems, the continuum of incoherent e-h excitations exhibits a similar behavior to that observed in the 3D case.  However, the charge collective excitations in 2D differ from that in 3D in long-wave-length limit. The plasmon energy $\omega_{2D}$ vanishes at small in-plane momenta ${\bf q}$ as $\omega_{2D}\sim\sqrt{ q}$.\cite{stprl67}
Such steep dispersion ensures a minimal spectral weight of this collective excitation at low energies resulting in a small effect on the low-energy properties of materials.

Much more relevant for the low-energy phenomena might be a mode, called acoustic plasmon (AP) with characteristic sound-like dispersion ($\omega_{\rm AP}$$\sim$$ q$), predicted to exist in a two-component electron system.\cite{picjp56}
This mode consists of out-of-phase collective movement \cite{siness06} of carriers in different energy bands which cross the Fermi level with different Fermi velocities.
The relevance of this mode as "a boson" for the superconductivity was discussed from the 60's.\cite{rapl65,gejetp65,frjpc68,ruleprb16,ruleprb17}
After discovery of superconductivity in cuprates, a role of the AP  as a possible candidate for mediation of the attractive interaction for the formation of Cooper pairs in high-T$_c$
superconductors was discussed.\cite{ruprb87,grprb88,krmoprb88,bimoprb02,bimoprb03,mahaprb08}

Existence of the AP in 3D bulk materials was predicted by a number of first principle response calculations.\cite{sibaprb09,sichprb09,kascprb13,zusiprb13,ecchprb19,cuwiprb21,gadinjp21}
However, it has never been observed in 3D bulk metals until recently.
Perhaps a signature of this mode was observed in intercalated graphite.\cite{rokoepl13}  But its measured energy dispersion was very different from the theoretical prediction\cite{ecchprb12} and therefore this issue still remains unclear.
In a recent publication\cite{huhun23} a claim on its experimental detection in a bulk sample of Sr$_2$RuO$_4$ was reported.  However, its measured dispersion presents qualitatively different behaviour from the predicted sound-like one. Therefore, to our opinion, this issue is still far from being settled.

\begin{figure}[h]%
\centering
\includegraphics[width=0.49\textwidth]{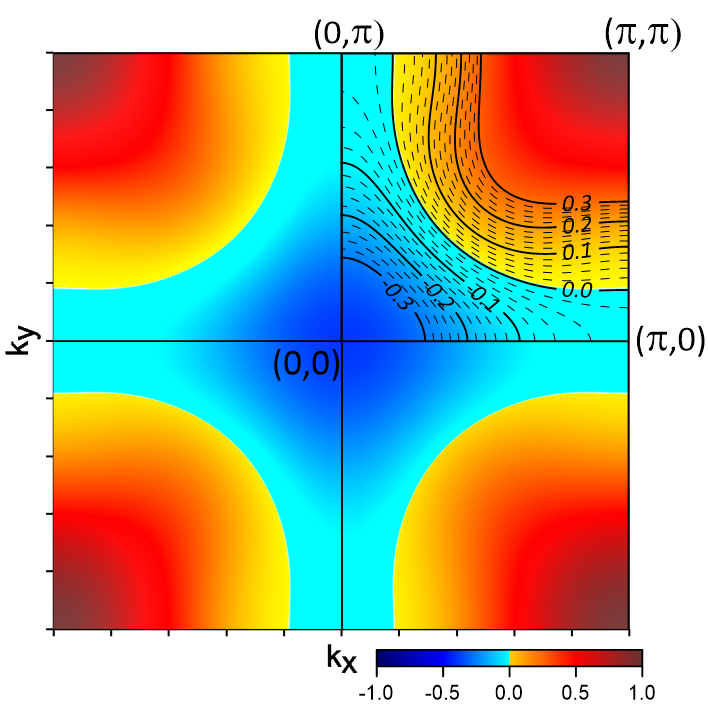}
\caption{
	 Electron dispersion in the optimally doped Bi-2212 as parameterized in the model of Ref.  \onlinecite{noraprb95}.
	The energy values are in eV. The Fermi level is set to zero.
In the upper-right corner some contour levels are highlighted by solid and dashed lines. }
\label{FIG-BS}
\end{figure}

On the other hand,  a cousin of the AP - an acoustic surface plasmon - which can be realized at metal surfaces supporting 2D-like electronic surface states\cite{sigaepl04} was detected on some surfaces like Be(0001),\cite{dipon07,jamuprb12}
noble\cite{papaprl10,vasmprl13,piwejpcc14,brvajpcl21} and transition\cite{bebesr21} metal (111) surfaces, supported graphene,\cite{labanjp10,tepfjpcm11,pomaprb11,lafonjp11,cuposs15} and topological insulators.\cite{posiprl15,gilaaom18,diadprl20,walajo20,povinc22,hawaap23}
Moreover, plasmons with a sound-like dispersion can be realized in MXenes\cite{papaam24} and in the atomically thin films where the quantum-well states can exist.\cite{pochpss15,baesnl23}

Usually, it is implicitly assumed that a multi-component electronic system can be realized when several energy bands cross the Fermi level. An example is a 2D dilute electron gas.\cite{anformp82}
Nevertheless recently it was demonstrated\cite{musipccp22} that an acoustic mode can exist, in addition to the conventional 2D plasmon, in a 2D system having a {\it single} energy band with a dispersion strongly deviating from the FEG isotropic case.
Thus, the effect of the energy dispersion anisotropy may not be limited to the changing only of the conventional 2D plasmon dispersion,\cite{haweprm21}  but can lead to the appearance at lower energies of additional modes in certain symmetry directions.

Cuprates are among the most promising materials for the observation of 2D plasmons, due to their quasi-2D electron band structure. 
It has been put forward that such plasmons can exist in these materials at energies below 1 eV.\cite{nuroprb89,ronuzpb90,nuecprb91,grpaprb99,rokojesrp14} Furthermore, it was suggested that  modifications of the plasmonic structure may be caused by a variation in the number of CuO$_2$ atomic sheets in the unit cell.\cite{grpiprb89,beziprb24}
Also, due to the layered structure of these materials the inter-layer long-range Coulomb interaction may transform the 2D plasmon mode into a gapped plasmon band. \cite{grprb73,feap74,krmoprb88,isruprb93,bimoprb03,gryaprb16}
Recently this mode was detected in the resonant inelastic x-ray scattering (RIXS) experiments.\cite{hechn18,liyuqm20,nazhprl20,sihuprb22,hebeprl22,heboprb23}

In a recent paper\cite{sidrjpcl23} it was predicted that in optimally doped cuprates characterized by a band dispersion similar to that presented in Fig. \ref{FIG-BS}, instead of substantial part of incoherent intra-band electron-hole pairs,
 unusual charge collective excitations  can exist. Namely taking as an example Bi-2212, it was demonstrated that the peculiar shape of the conducting energy band  gives rise to new kinds of modes with unusually strong spectral weight called hyperplasmons of types I and II.
Additionally, an uni-directional mode becoming soft along the ${\bf q}$$=$$(q_x,0)$ and ${\bf q}$$=$$(0,q_y)$ lines (with $q_x,q_y$ varying from 0 to $\approx0.4\pi$) in the ${\bf q}$ space was found. All these modes can be realized in the energy interval from zero to several hundreds of meV.

Let us stress that the properties of Bi-2212 are crucially dependent on the doping level. Therefore, if the electronic modes discovered in Ref. \onlinecite{sidrjpcl23} have any implication, it would be interesting to reveal how the modes properties change upon the doping level.
Moreover, the dispersion of these modes is very close to that of the paramagnon at small momenta. Actually it falls into broad experimental features assigned to paramagnon.\cite{leghnp11,dedenm13,lemiprb13,isfanc14,gupinc14,lelenp14,memiprb17,ivshprb17,demiprb17,chhuprb18,fubrprb19,robaprb19,zhagqm22,lituprl24}
We believe that thorough doping dependent experimental study  might disentangle the paramagnon and the electronic modes of interest here.

In this paper we analyze how variation in the electronic structure at the Fermi level caused by doping variation are reflected in the intra-band electronic excitations. In particular, we show that the balance between the intra-band e-h's and the collective excitations is very sensitive to the doping level. We relay this strong sensitivity to the fast variations in the Fermi line shape and the Fermi velocity distribution caused by doping.

\section{Calculation Details}\label{Detailes}

In order to obtain the response function $\chi^o({\bf q},\omega)$ as a function of momentum ${\bf q}$ and energy transfer $\omega$ for non-interacting electrons we calculate the spectral function $S^o({\bf q},\omega)$ as
\begin{equation}
S^o({\bf q},\omega)= \frac{2}{A} \sum_{\bf k}^{BZ} (f_{\bf k}-f_{{\bf k}+{\bf q}}) \delta(\varepsilon_{\bf k}-\varepsilon_{{\bf k}+{\bf q}}+\omega).
\label{S0}
\end{equation}
Here the factor 2 accounts for the spin, $A$ is the unit cell size, the Fermi distribution function $f_{\bf k}$ is a step function, and $\varepsilon_{\bf k}$ is the conducting one-particle band dispersion. As in Ref. \onlinecite{sidrjpcl23}, in Eq. (\ref{S0}) a mesh of $2500\times2500$ ${\bf k}$ points was employed in the summation over the two-dimensional Brillouin zone (2DBZ).
Then the imaginary part of  the density-density response function $\chi^o({\bf q},\omega)$ is evaluated according to
\begin{equation}
{\rm Im}[\chi^o({\bf q},\omega)] = -sgn(\omega)\pi S^o({\bf q},\omega)
\end{equation}
and the real part of $\chi^o({\bf q},\omega)$ is evaluated via the Kramers-Kronig relation. Finally, the dielectric function $\epsilon({\bf q},\omega)$ is obtained as
\begin{equation}
\epsilon({\bf q},\omega) \equiv \epsilon_1({\bf q},\omega)+{\rm i}\epsilon_2({\bf q},\omega)=\epsilon_{\infty}-V({\bf q})\chi^o({\bf q},\omega).
\end{equation}
Here $\epsilon_1$ and $\epsilon_2$ are the real and imaginary parts of the dielectric function, respectively. The experimental value for the background dielectric constant $\epsilon_{\infty}$ is 4.5.\cite{letrprx16} Nevertheless, we have checked that the exact $\epsilon_{\infty}$ value does not change our results. Therefore, in the following we use $\epsilon_{\infty}=1$.
Previously, similar random phase approximation (RPA) calculations were employed in the study of plasmons in cuprates.\cite{krmoprb88,bimoprb03,mahaprb08,papeltp08,sihuprb22,beziprb24}
For the interacting potential we employ a 2D Coulomb potential, $V({\bf q})=2\pi/q$. Actually, the shape of the interacting potential should not influence notably the properties of the plasmon modes studied here. These modes correspond to the out-of-phase charge oscillations at different parts of the Fermi surface,\cite{sidrjpcl23} whereas in the conventional 2D plasmon all the electronic system oscillates in phase. Therefore the interlayer Coulomb interaction is more efficient in the latter case.\cite{besiprb03}

\begin{table}
\centering
\begin{center}
\begin{tabular}{|c|c|c|}
	\hline
 $i$ &	$c_i$ (in eV)&  $\eta_i(\mathbf{k})$  \\
	\hline
 $0$	& 0.1305&  1 \\
 $1$	&	-0.5951& $\frac{1}{2}(\cos k_x + \cos k_y)$ \\
 $2$	&	0.1636 & $\cos k_x  \cos k_y $  \\
 $3$	&	-0.0519& $\frac{1}{2}(\cos 2k_x + \cos 2k_y) $ \\
 $4$	&	-0.1117 &$ \frac{1}{2}(\cos2 k_x \cos k_y + \cos k_x \cos 2 k_y)$ \\
 $5$	&	0.0510 &  $\cos2 k_x  \cos 2k_y $\\
	\hline
\end{tabular}
\end{center}
	\caption{ Parameters of the tight-binding model describing the band structure in the normal state of Bi-2212 at the optimal doping, Eq. (\ref{TB_mod}), as proposed in Ref. \onlinecite{noraprb95}.}
\label{tab.parameters_of_TB}.
\end{table}

\begin{figure}[ht]%
\centering
\includegraphics[width=0.48\textwidth]{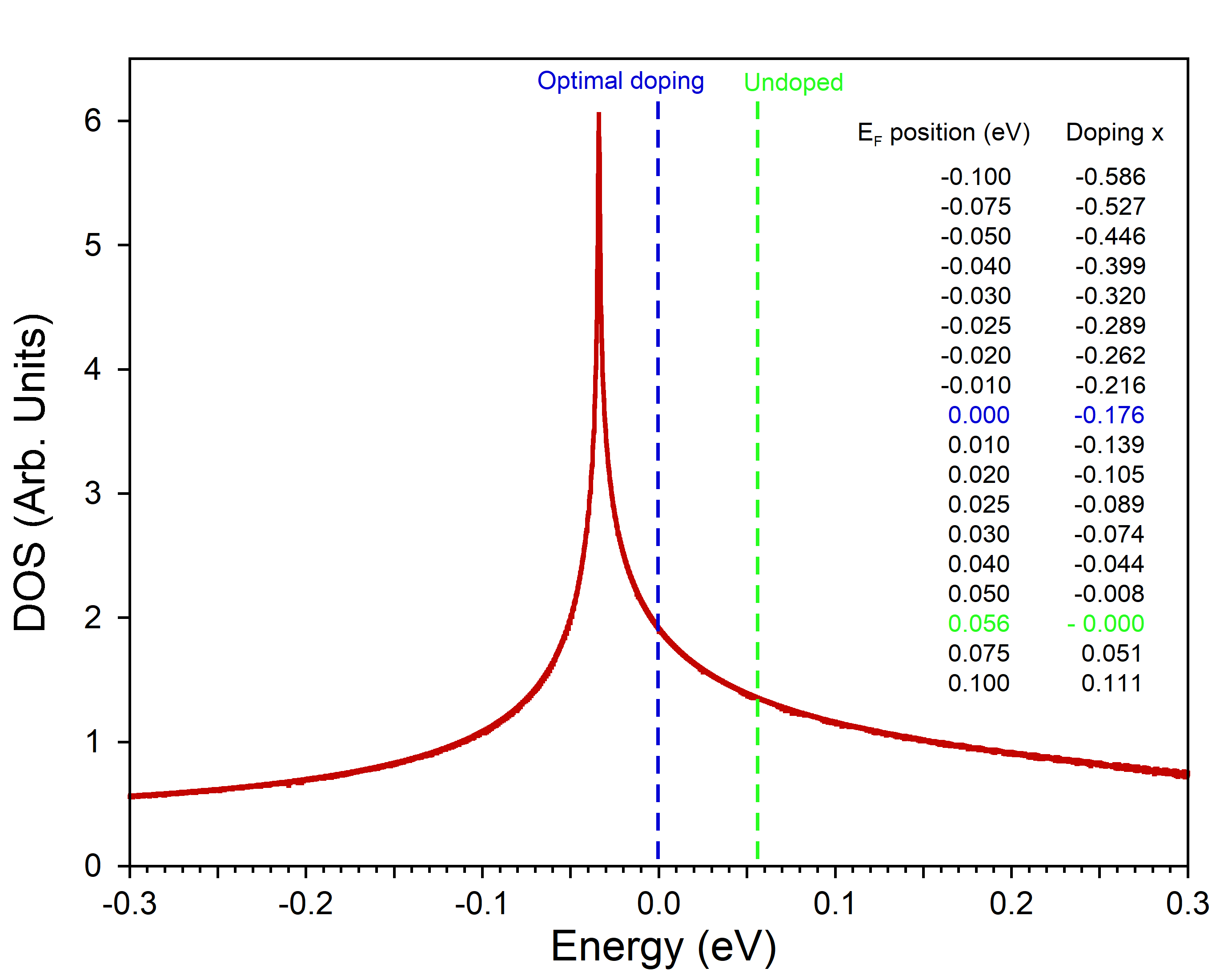}
\caption{Density of states around the Fermi level in the metallic energy band in Bi-2122 as parameterized in Ref. \onlinecite{noraprb95}. The Fermi level is set to zero in the optimally doped case. In the insert the relation between certain Fermi level positions and the doping level is reported.
}
\label{FIG-DOS}
\end{figure}

\section{Results}

In this work we consider a 2D case of a free standing Bi-2212 monolayer. It is justified since we are interested in the collective electronic modes different from the conventional 2D plasmon and quasi-2D gapless plasmon band studied in details previously.

In a normal state of the optimally doped Bi-2212, the dispersion of the energy band $\varepsilon_{\bf k}$  crossing the Fermi level can be described in the tight-binding (TB) model\cite{noraprb95}

\begin{equation}
	\varepsilon_{\bf k} = \sum_i c_i \eta_i (\mathbf{k})
\label{TB_mod}
\end{equation}
with parameters $c_i$ and $\eta_i$  listed in Tab. \ref{tab.parameters_of_TB}. In the band structure plot reported in Fig. \ref{FIG-BS} one can see that the Fermi line has a characteristic shape with straight regions in the vicinity of the nodal points [$(\pm \pi,0)$ and $(0,\pm \pi)$]. Moreover  in the Fermi level vicinity, as seen in the upper-right part of Fig. \ref{FIG-BS}, the shape of the lines of constant energies (LCE) changes quickly as well. This is very different from the FEG model, where all the LCEs are the circles.
The difference between the realistic Bi-2122 band dispersion of Fig. \ref{FIG-BS} and the 2D FEG one is seen in the peak structure of the density of states (DOS) in the former case (reported in Fig. \ref{FIG-DOS}) instead of a step function in the latter one.

We start with analyzing the dynamic part of the density-density response function at $T=0$.
For simplicity we use the system of coordinates centered at $(\pi,\pi)$.
For small $q\ll p_F$ we use the expansion of the quasiparticle spectra $\varepsilon_{\mathbf{p+q}} \approx \varepsilon_\mathbf{p} + \mathbf{v}_F(\varphi) \mathbf{q}$.
Integration over $\varepsilon_\mathbf{p}$ followed by $\omega$ gives the dynamic part of the response function

\begin{equation}
\tilde\chi^o(\mathbf{q},i \omega) = N(0)\int \frac{d\varphi }{2\pi}\frac{i \omega}{i\omega + \mathbf{v}_{F}(\varphi)\mathbf{q}} ,
\label{chi0_tilde}
\end{equation}	
where $N(0)$ is the density of states at the Fermi level.

\begin{figure}
	\centering
	\includegraphics[width=0.95\linewidth]{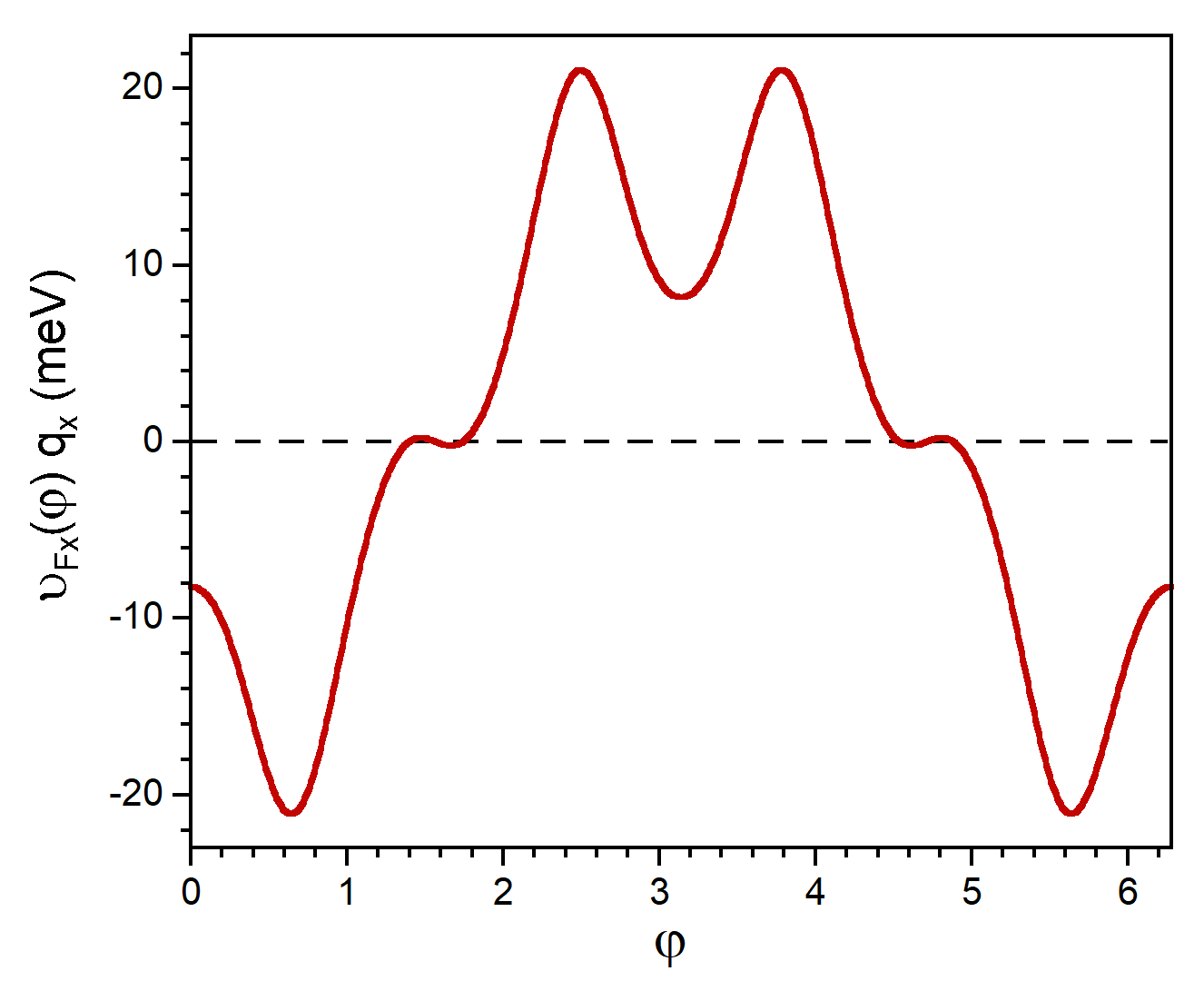}
	\caption{The $x$ component of ${\bf v}_F$ multiplied by $q_x$ at $q_x = 0.02\pi$ as a function of angle for the Fermi surface centered at $(\pi,\pi)$.}
	\label{fig:velocityx}
\end{figure}

\begin{figure}
	\centering
	\includegraphics[width=0.95\linewidth]{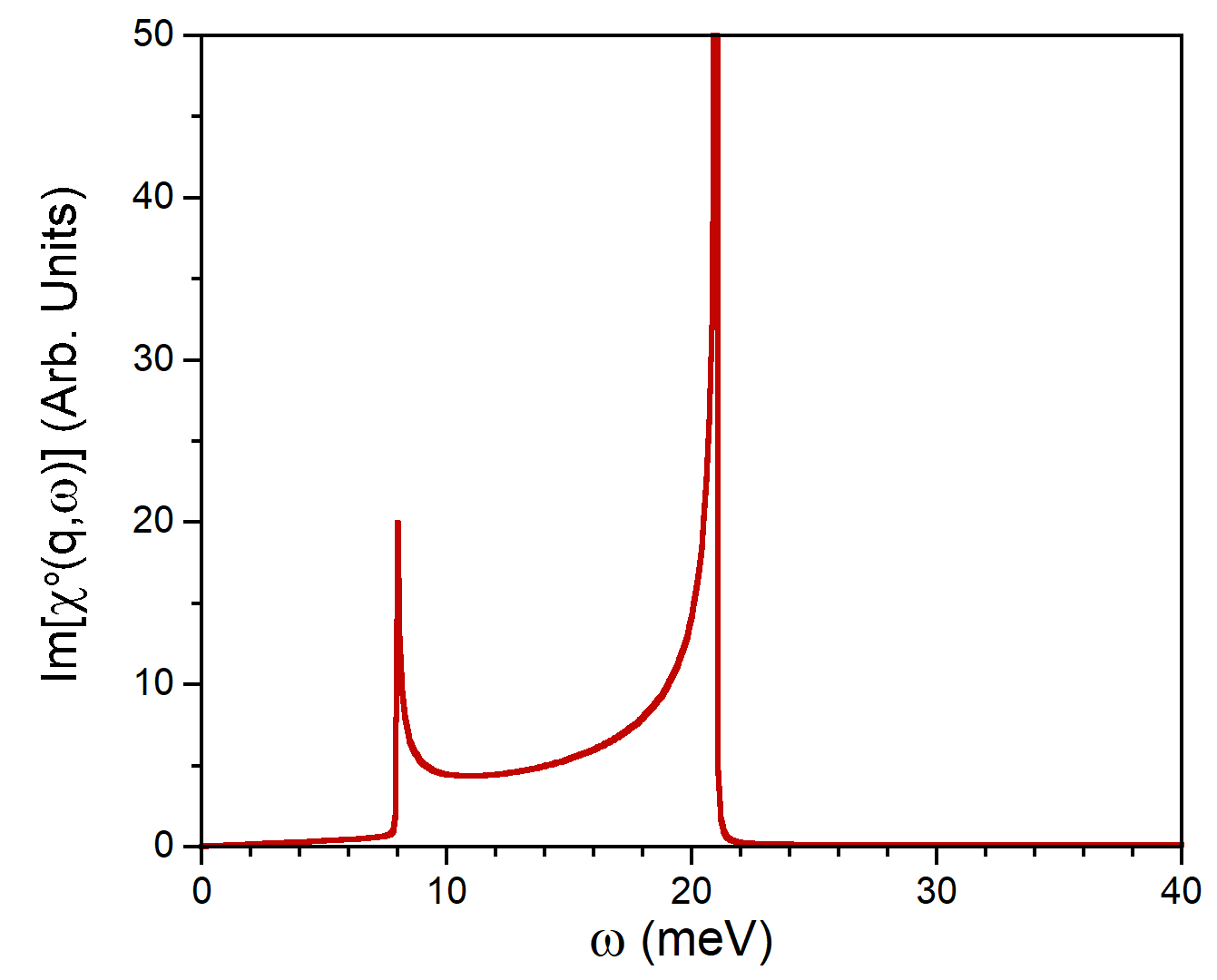}
	\caption{Im[$\chi^o(\mathbf{q},\omega)$] at $\mathbf{q} = (0.02\pi,0)$ as a  function of $\omega$.}
	\label{fig:chi0}
\end{figure}

For the chosen dispersion we can write out $\mathbf{v}_{F}(\varphi)\mathbf{q}  =|\mathbf{v}_{F0}|q\cos(\varphi)(1 + a_3 \cos^2(\varphi)+a_5 \cos^4(\varphi))$, where  $|\mathbf{v}_{F0}|$, $a_3$ and $a_5$ are some coefficients.
The corresponding function is shown in Fig. \ref{fig:velocityx}. It can be seen that the angular dependence of the Fermi velocity can be divided into three intervals. The first interval contains a double peak angle dependence centered at $\varphi=\pi$, the second is a double dip near $\varphi=0$, and the last one are inflection points at $\varphi=\pi/2, 3\pi/2$.
The Fermi velocity in the double peak interval can be approximated as
$\mathbf{v}_{F}(\varphi)\mathbf{q} \approx \omega_2 - \frac{\omega_2-\omega_1}{\varphi_0^4} (\varphi^2-\varphi_0^2)^2$, where $\omega_{2}$ and $\omega_{1}$  are the maximum and the minimum of $\mathbf{v}_{F}(\varphi)\mathbf{q}$ in this interval,
 and $\varphi_0$ is the maximum position. As the integral converges rapidly, it is possible to extend the limits of integration to infinity. Following the integration, the function will assume the following form
\begin{eqnarray}
	\tilde\chi^o(\mathbf{q},\omega) &\approx&- \frac{i N(0)}{2\sqrt{2}} \frac{\varphi_0 }{(\Delta \omega)^{1/4}}
	\nonumber \\ &&\times \frac{\omega\sqrt{(\omega -\omega_{1} )^{1/2}+(\Delta \omega)^{1/2}}}{\sqrt{\omega-\omega_1} \sqrt{\omega_{2}-\omega}},
\label{eq.icmchi1}
\end{eqnarray}
where $\Delta \omega =\omega_{2}-\omega_{1} $.
The double dip interval gives similar result with opposite sign of $\omega_{1,2}$.
The intervals with the inflection points can be estimated as
\begin{eqnarray}
	\tilde\chi^o(\mathbf{q},\omega) &\approx&
	2N(0)\int_{-\infty}^{\infty} \frac{d\varphi }{2\pi}\frac{ \omega}{\omega + a \varphi^3 q+i0 }
	 \nonumber \\ &&=
	N(0) \frac{\omega^{1/3}}{3(aq)^{1/3}}\left(\sqrt{3} - i\right).
	\label{eq.icmchi2}	
\end{eqnarray}
The imaginary part of the response function is presented in Fig. \ref{fig:chi0}.
The two peaks in the imaginary part of the response function originate from the term in the denominator  $\sqrt{(\omega-\omega_{1})(\omega_{2}-\omega)}$ of Eq. (\ref{eq.icmchi1}). The contribution from the inflection points, which is $\sim \omega^{1/3}$, does not result in the singular inverse power law contribution to the response function.

In 2D systems, the inverse square root singularity is a typical characteristic of response functions.  For an isotropic Fermi-gas with parabolic dispersion it has a form of $1/\sqrt{\omega^2 - (\mathbf{v}_{F} \mathbf{q})^2}$.
In the case of a nonmonotonic dependence of $\mathbf{v}_F$ for the electron band structure, as occurs in high-T$_c$ cuprates, the extremums of $\mathbf{v}_F\mathbf{q}$ yield an inverse square root singularity in the response function, as described in Eq.  (\ref{eq.icmchi1}).

\begin{figure}[ht]%
\centering
\includegraphics[width=0.48\textwidth]{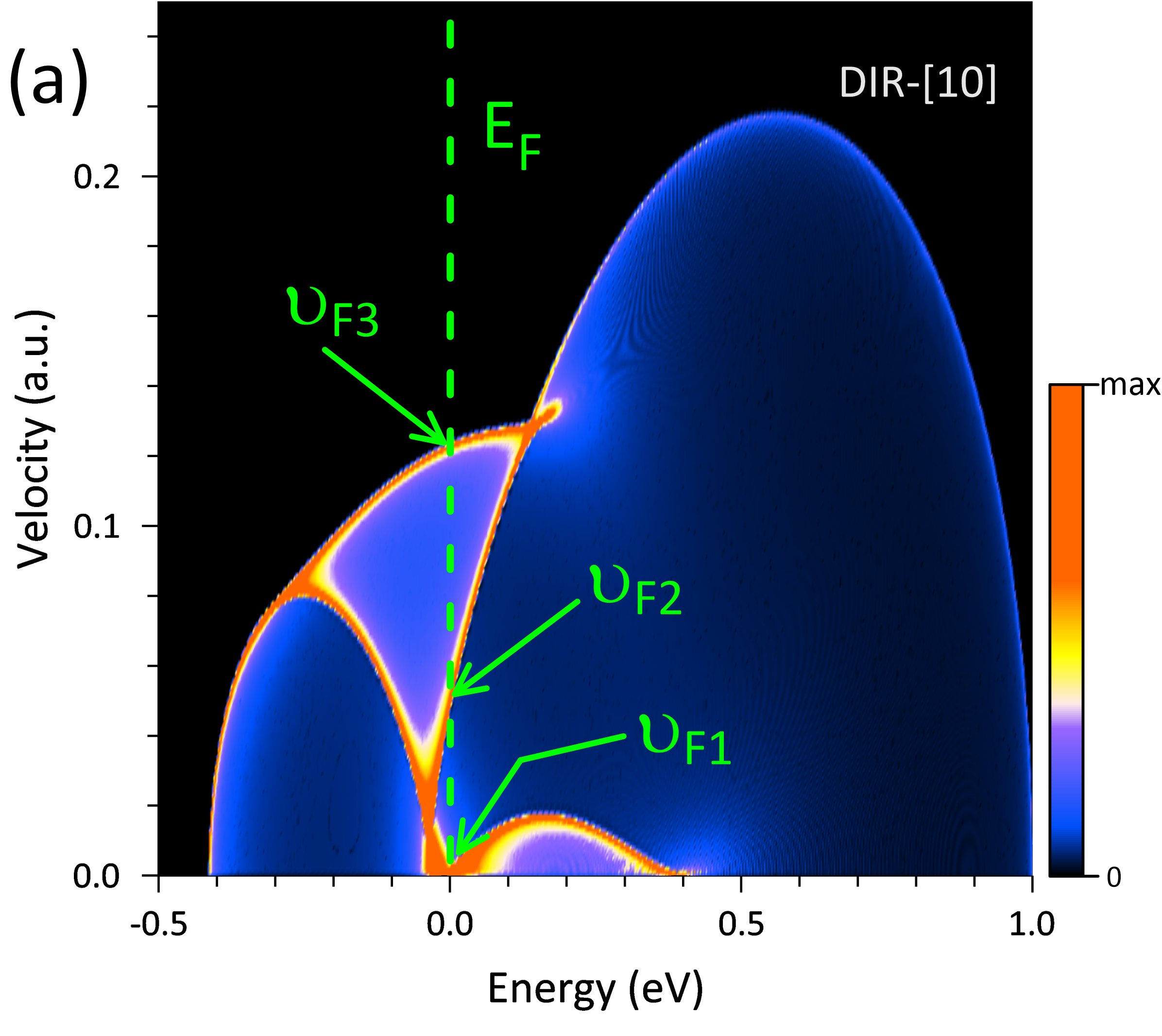}
\includegraphics[width=0.48\textwidth]{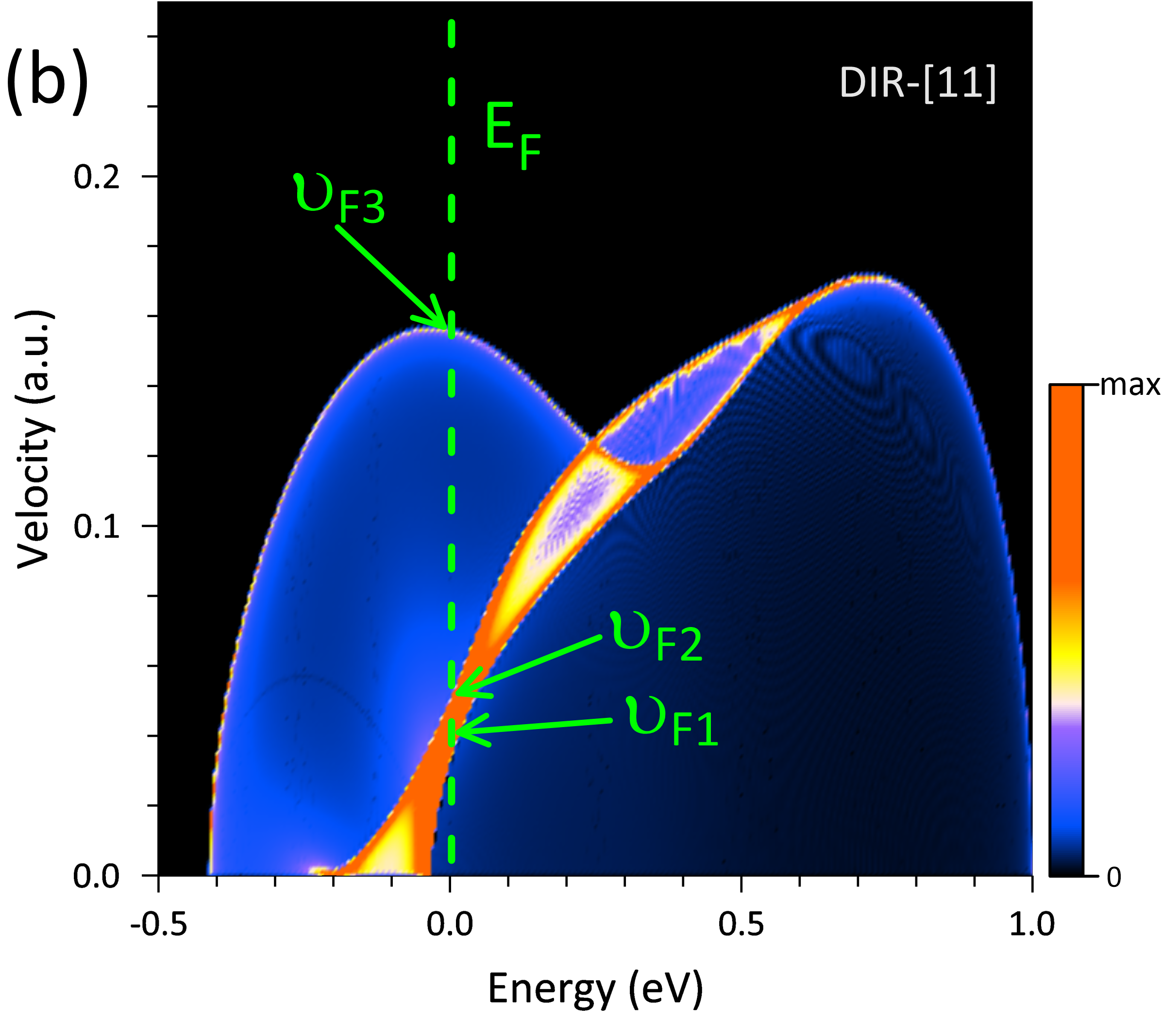}
\caption{ Two-dimensional plot of the density of states versus energy and group velocity component along (a) [10] and (b) [11] symmetry directions.  The Fermi level is set to zero energy. The peaks in the DOS at the Fermi level are marked by respective Fermi velocity values, $v_{F1}$, $v_{F2}$, and $v_{F3}$.
 }
\label{FIG-GV}
\end{figure}

\begin{figure}[ht]%
\centering
\includegraphics[width=0.48\textwidth]{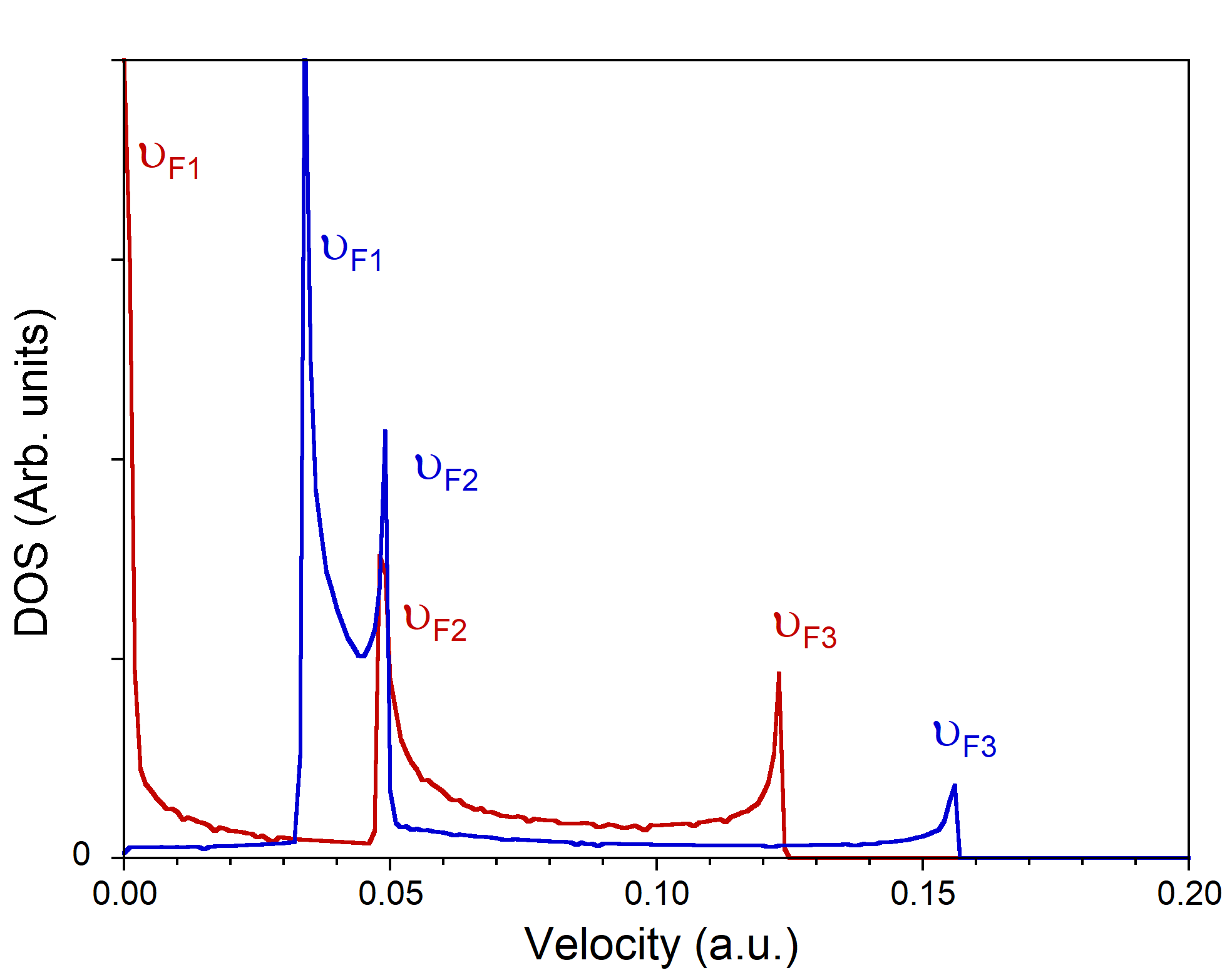}
\caption{Density of states versus group velocity of the carriers moving at the Fermi level in the [10] (red line) and [11] (blue line) symmetry directions. Optimally doped case.
}
\label{DOS_vs_GV_at_EF}
\end{figure}

After establishing the general properties of the response function in the limit of small q, we proceeded to calculate the doping and momentum dependence of the plasmon dispersion using a numerical approach.
 Quick variations in the Bi-2122  band dispersion are reflected in the non-isotropic distribution of the group velocities. An useful quantity to analyze is the distribution of the density of states DOS($E,v$)
 as function of an electron energy $E$ and a group velocity component $v$ in a certain direction.
 The DOS for the group velocity components $v_{[10]}$ and $v_{[11]}$ along the [10] and [11] symmetry directions, respectively,  is  reported in Fig. \ref{FIG-GV}. For more details, these quantities at the Fermi level (set at zero energy in the optimally doped case, $x$=-0.176), DOS($E_F,v_{[10]}$) and DOS($E_F,v_{[11]}$), are presented in Fig. \ref{DOS_vs_GV_at_EF}. One can see that the DOS for the carriers moving at the Fermi level in the [10] direction is characterized by three sharp peaks at $v_{F1}$=0, $v_{F2}$=0.049, and $v_{F3}$=0.123 a. u. Along the [11] direction the carriers are moving at the Fermi level essentially with velocities $v_{F1}$=0.034, $v_{F2}$=0.049, and $v_{F3}$=0.156 a. u. Such distribution of the peaks in the  DOS is very different from that in the FEG model, where a single peak locates at $v_F$ in any direction.\cite{pino66,givi08}

\begin{figure}[ht]%
\centering
\includegraphics[width=0.45\textwidth]{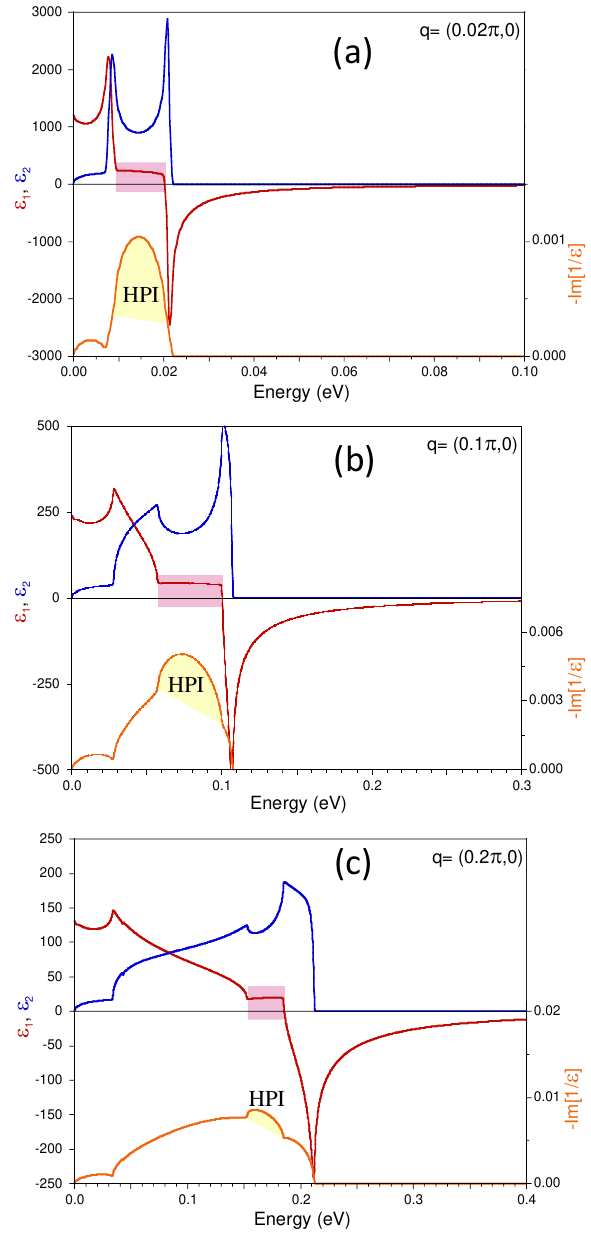}
\caption{Dielectric and loss functions obtained with the tight-binding energy band dispersion of Ref. \onlinecite{noraprb95} at (a) ${\bf q}=(0.02\pi,0)$, (b) ${\bf q}=(0.1\pi,0)$, and (c) ${\bf q}=(0.2\pi,0)$,  along the [10] symmetry  direction.
The peaks corresponding to the hyperplasmon of type I are marked as HPI. The colored boxes highlight the region where $\epsilon_1$ has a horizontal line dispersion.
}
\label{EPS-TB_FULL_DIR-10}
\end{figure}

In Ref. \onlinecite{sidrjpcl23} the presence of three charge collective excitations resulted from such peak structure of DOS($E,v$) was linked to the different regions in the vicinity of $E_F$ characterized by different Fermi velocities. This is because at small momentum transfers ${\bf q}$'s, the intra-band contribution to the imaginary part of the dielectric function is essentially determined by the peak structure in DOS($E_F,v$).\cite{pino66,givi08}
As a result, a multi-peak behavior of the DOS$(E_F,v)$ results in a multi-peak structure of the imaginary part of the dielectric function, $\epsilon_2(\bf q,\omega)$. Thus, at momentum transfers along the [10] direction two clear peaks can be observed in $\epsilon_2$, especially at small momentum transfers ${\bf q}$'s, as it is seen in Fig. \ref{EPS-TB_FULL_DIR-10}.
 In turn, the real part $\epsilon_1$, which is connected to $\epsilon_2$ by the Kramers-Kroning relation, has unconventional behavior as well. Its remarkable feature is the near horizontal dispersion over an extended energy regions for ${\bf q}$'s along [10]. In Fig. \ref{EPS-TB_FULL_DIR-10} these regions are highlighted by colored rectangles. For instance,  in Fig. \ref{EPS-TB_FULL_DIR-10}(a) such an energy region with flat $\epsilon_1$ expands from $\omega$=9.5 meV to $\omega$=19.5 meV. In Fig. \ref{EPS-TB_FULL_DIR-10}(b) a similar interval is between $\omega$=57 meV and $\omega$=99 meV. In the case of ${\bf q}=(0.2\pi,0)$, presented in Fig. \ref{EPS-TB_FULL_DIR-10}(c), such a region in $\epsilon_1$ locates between $\omega$=153 meV and $\omega$=185 meV.  Moreover, in these flat regions a value of $\epsilon_1$ is significantly smaller than  $\epsilon_2$.
 As a consequence, the imaginary part of the dielectric function almost completely determines\cite{sidrjpcl23} the imaginary part of the inverse dielectric function, i.e. the loss function which is directly measured in the loss experiments.\cite{pino66}
 Taking into consideration that in these energy intervals (a) $\epsilon_1 \ll \epsilon_2$, (b) $\epsilon_2$ has shallow minima, and (c) the loss function -Im$[1/\epsilon]$ presents a peak, the spectral weight of which (marked in yellow) in the loss functions of Figs. \ref{EPS-TB_FULL_DIR-10}(a,b,c) was attributed to a novel collective charge excitation called hyperplasmon of type I (HPI). Notice, that the spectral weight of the HPI is notably larger in comparison to the usual acoustic plasmon appearing in a two-band scenario with the isotropic FEG energy band dispersion at comparable momenta.\cite{sidrjpcl23}

\begin{figure}[ht]%
\centering
\includegraphics[width=0.45\textwidth]{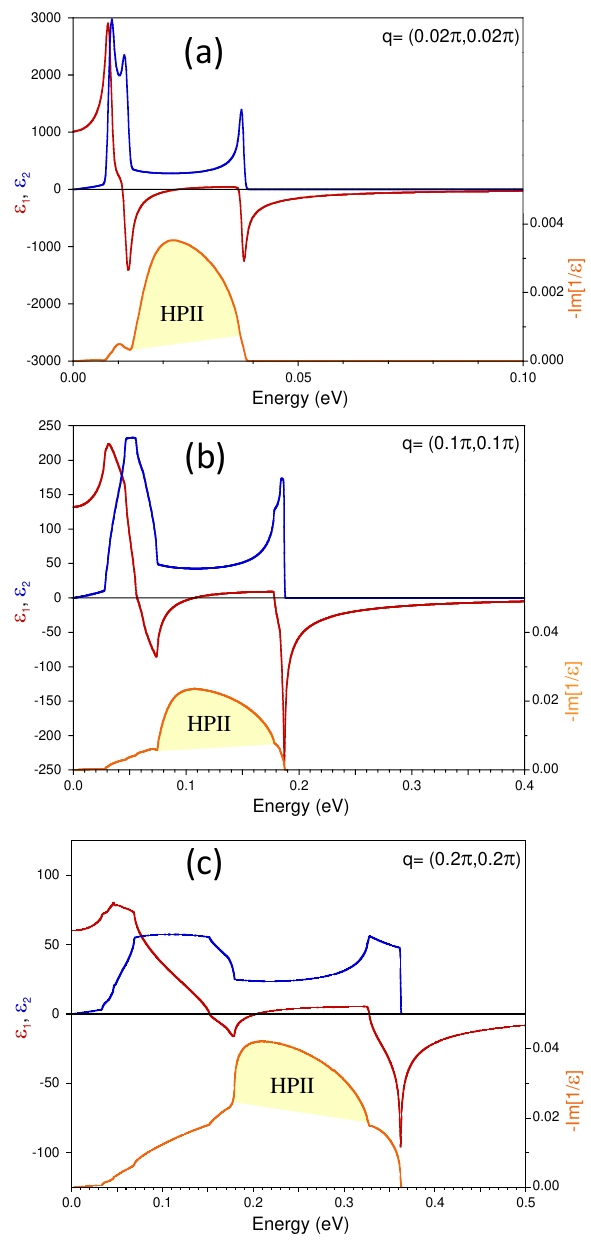}
\caption{Dielectric and loss functions obtained with the tight-binding energy band dispersion of Ref. \onlinecite{noraprb95} at (a) ${\bf q}=(0.02\pi,0.02\pi)$, (b) ${\bf q}=(0.1\pi,0.1\pi)$, and (c) ${\bf q}=(0.2\pi,0.2\pi)$,  along the [11] symmetry  direction.
The peaks corresponding to the hyperplasmon of type II are marked as HPII.
}
\label{EPS-TB_FULL_DIR-11}
\end{figure}

As seen in Fig. \ref{EPS-TB_FULL_DIR-11}, at ${\bf q}$'s along the antinodal [11] direction the amplitude of the real part of dielectric function again is significantly smaller than $\epsilon_2$ in the energy interval where a broad prominent peak highlighted by yellow color is observed in the loss function. However, in this case $\epsilon_1$ changes the sign in this energy interval. In consequence, the spectral weight of a resulting collective mode is significantly larger in comparison to the case of HPI. Therefore, the respective mode was called a hyperplasmon of type II (HPII).\cite{sidrjpcl23}

In the low-energy excitation spectrum of Bi-2122, besides the HPI and HPII modes, in Ref. \onlinecite{sidrjpcl23} another mode was found. Its peculiarity is the quasi-one-dimensional dispersion along the [10] and [01] symmetry directions. For this reason it was called
 a quasi one-dimensional plasmon (1DP). This mode has finite energies at ${\bf q}$'s in vicinity of the nodal [10] and [01] symmetry directions whereas it is a soft mode (i.e. has a zero energy) exactly along the lines from ${\bf q}=(0,0)$ to ${\bf q}=(\approx 0.4\pi,0)$ and ${\bf q}=(0,\approx 0.4\pi)$. Existence of this mode was explained by the flat regions on the Fermi lines occurring at the optimal doping in the vicinity of the nodal $(\pi,0)$ and $(0,\pi)$ points.\cite{sidrjpcl23}

 In Ref. \onlinecite{sidrjpcl23} all the unusual features in the low-energy excitation spectra of optimally doped Bi-2212 were linked to the remarkable dispersion of the metallic energy band and, especially, to the peak structure in the distribution of the Fermi velocities. Therefore a reasonable question arises - what would happen with these excitations upon variation of the doping level. From the band structure of Fig. \ref{FIG-BS} it is clear, that with shifting of the Fermi level the shape of the Fermi line changes notably. This is accompanied by quick change of the energies of peaks in the DOS of Fig. \ref{FIG-GV} as well. In this paper we address this question by shifting the Fermi level position, i.e. mimicking in such a way the variation in the doping level. After that, at each shifted Fermi level position, the response calculations are realized. The relation of the doping level with the Fermi level position is reported in the insert of Fig. \ref{FIG-DOS}.

\begin{figure*}[ht]%
\centering
\includegraphics[width=0.92\textwidth]{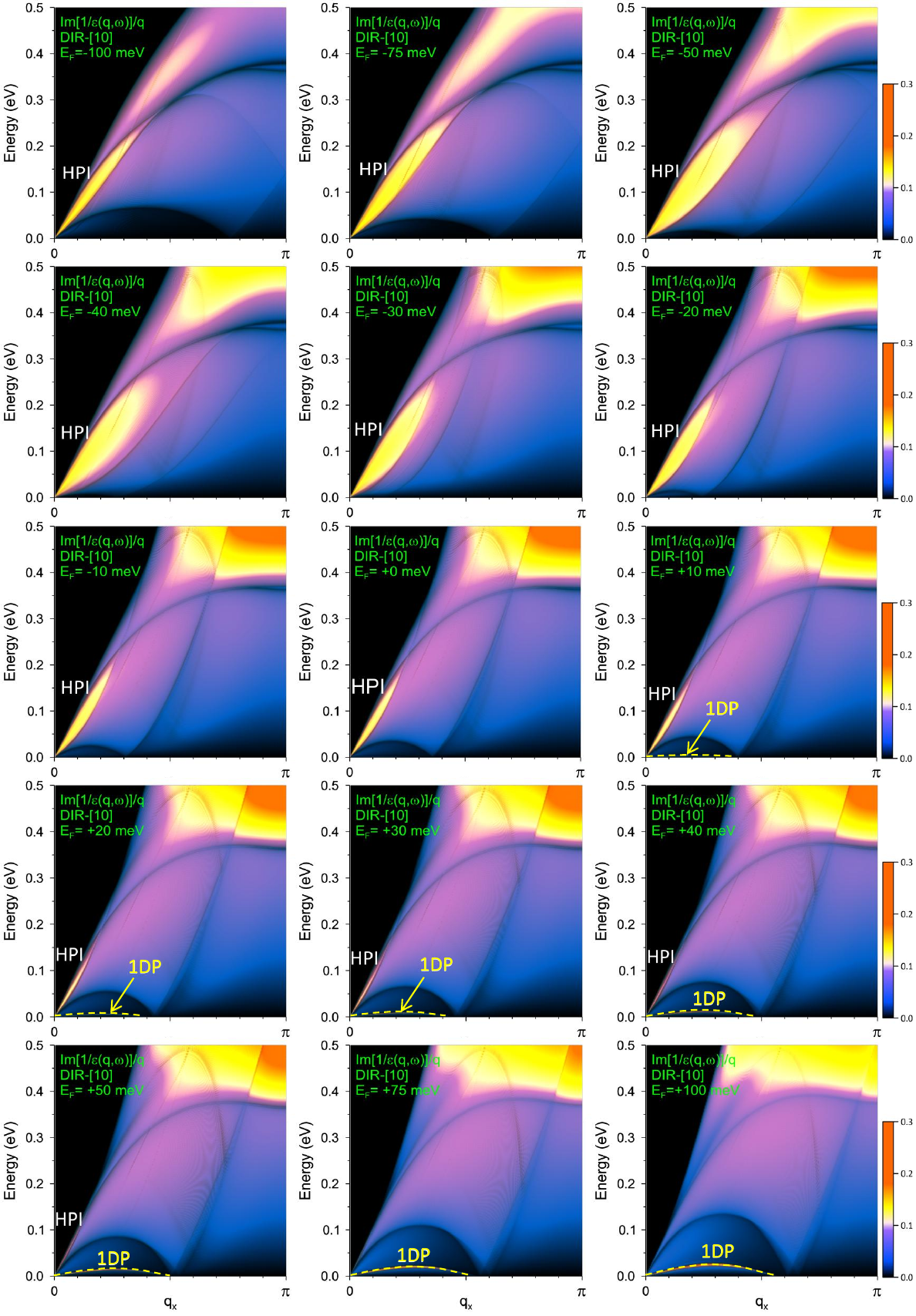}
\caption{ Normalized loss function $L({\bf q},\omega)$=-Im$[1/\epsilon({\bf q},\omega)]/\omega$ at momentum transfers ${\bf q}$ along the $[10]$ direction.  Different panels represent $L({\bf q},\omega)$ for the Fermi level positions respective to its optimally doping position. The peaks corresponding to the hyperplasmon of type I and the quasi one-dimensional plasmon are marked by HPI and 1DP, respectively.
The dispersion of the conventional two-dimensional plasmon 2DP at small ${\bf q}$'s outside the electron-hole continuum is not shown.
}
\label{LOSS-2D_DIR-10}
\end{figure*}


In Fig. \ref{LOSS-2D_DIR-10} the normalized loss function at momentum transfers along the [10] symmetry direction at different $E_F$ position (a respective value is reported in each panel) is presented. We vary the $E_F$ from -0.1 eV to 0.1 eV according to its optimal doping position.
At the optimal doping ($E_F$=0) the HPI peak is seen in the lower-left corner of the panel. Upon shifting  $E_F$ downward the intensity of this peak increases. Also the HPI peaks becomes wider. The peak width reaches a maximal value at around $E_F$=-40 meV. Upon subsequent lowering of the Fermi level positions it reduces gradually again.

\begin{figure}[ht]%
\centering
\includegraphics[width=0.48\textwidth]{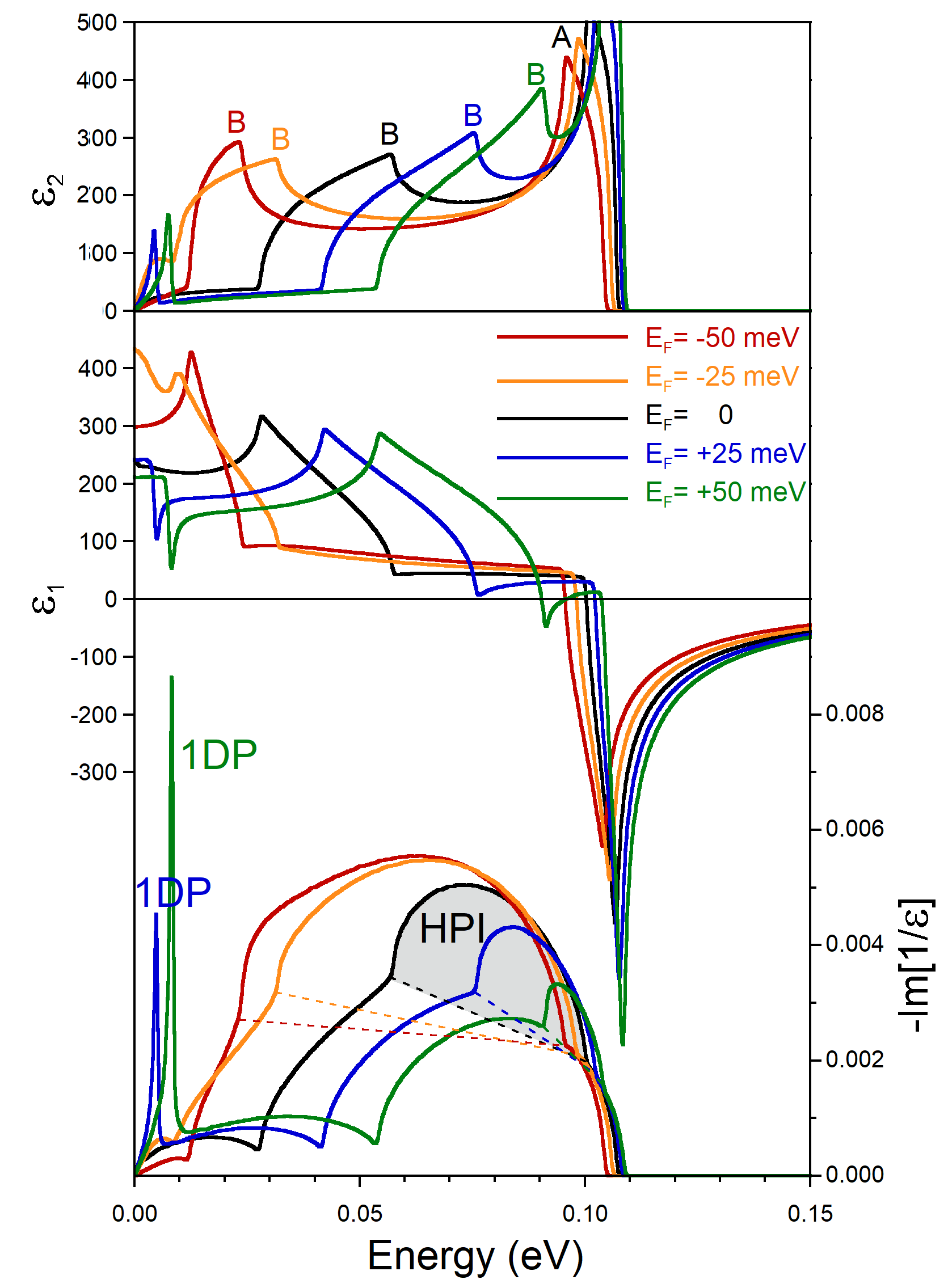}
\caption{ Imaginary (top panel) and real (middle panel) parts of the dielectric function at momentum transfer ${\bf q}=(0,1\pi,0)$ for five Fermi-level positions.  The respective loss functions are presented in the bottom panel. Spectral weight in the loss function at $E_F=0$ tentatively corresponding to the hyperplasmon HPI is highlighted by grey color. Thin dashed lines delimit such regions for all five doping levels. The peaks in the loss function at $E_F$=+25 and +50 meV corresponding to the quasi-one dimensional plasmon are marked as 1DP.
}
\label{EPS_Q=0250_0000}
\end{figure}

In order to demonstrate how the HPI spectral weight varies with doping, in Fig. \ref{EPS_Q=0250_0000} we report the real and imaginary parts of dielectric function as well as the loss function calculated at ${\bf q}=(0.1\pi,0)$ for five values of the doping level.
One can see that, upon downward shift of $E_F$, the energy of the right peak A in $\epsilon_2$ slightly reduces, which correlates with a small reduction of the $v_{F3}$ value on the left side of the Fermi level position in Fig. \ref{FIG-GV}(a). On the contrary, the left peak B in $\epsilon_2$ shifts downward dramatically. This can be explained by strong reduction of $v_{F2}$ from its value of 0.049 a.u. at the Fermi level to a zero value at about $E=-50$ meV, as observed in Fig. \ref{FIG-GV}(a).
As a result, upon the lowering of the Fermi level position the energy interval between two peaks in $\epsilon_2$ increases.
 This results in increasing of the energy interval where $\epsilon_1$ has a near straight line dispersion (being at the same time notably smaller than $\epsilon_2$) as well. For instance, in the case of $E_F=-50$ meV, this energy interval spans a region from $\omega\approx$25 to $\omega\approx$95 meV. At the same time, upon the downward shift of $E_F$ the HPI-peak width in the loss function increases as well. Also the spectral weight of this peak in the loss function increases (compare the regions in the loss functions above the orange and red thin dashed lines with that shaded in grey color) as well as the total spectral weight. Such increase of the total spectral weight correlates with the increase of the doping level.

\begin{figure*}[ht]
\centering
\includegraphics[width=0.92\textwidth]{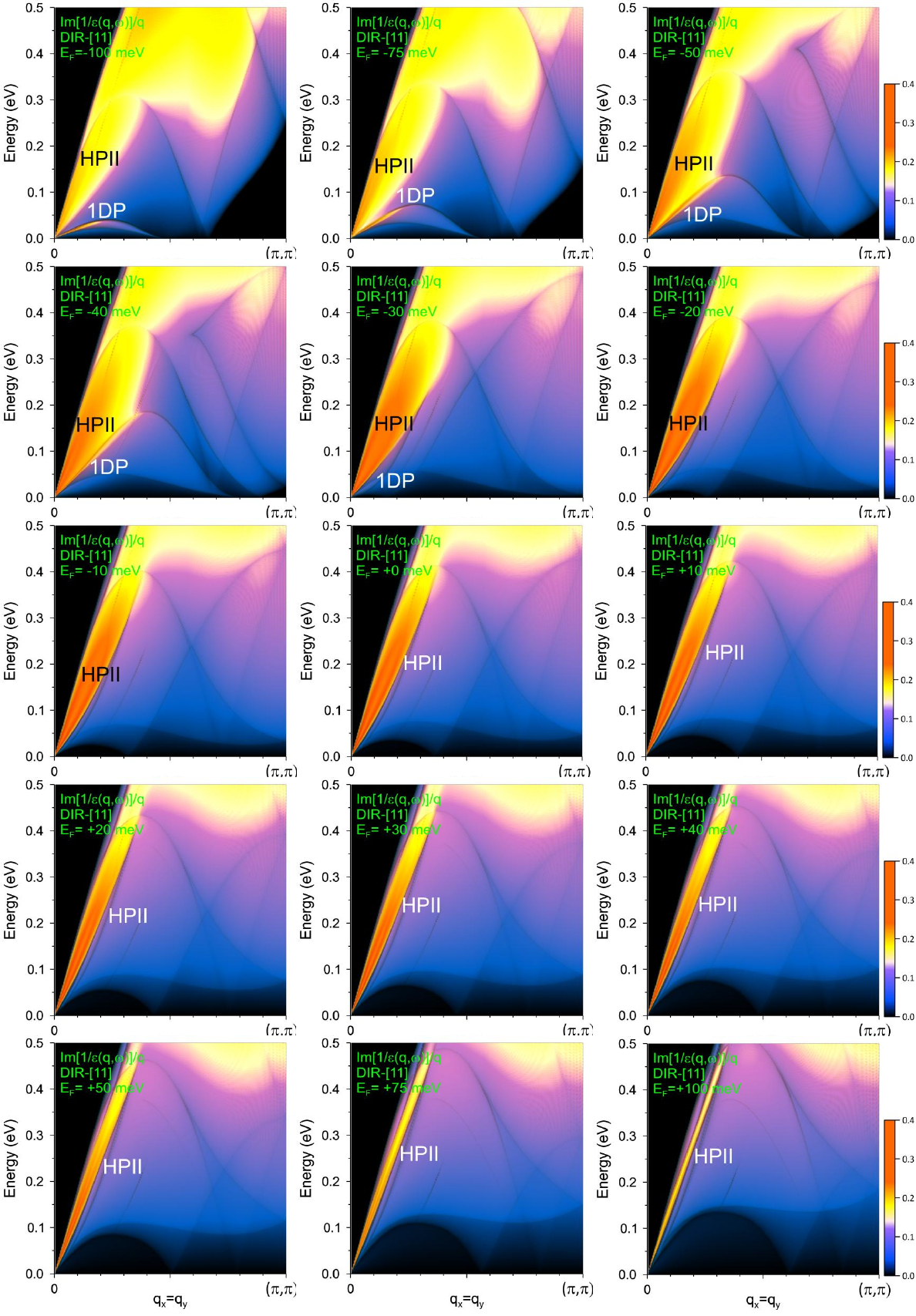}
\caption{ Normalized loss function $L({\bf q},\omega)$=-Im$[1/\epsilon({\bf q},\omega)]/\omega$ at momentum transfers ${\bf q}$ along the $[11]$ direction.  Different panels represent $L({\bf q},\omega)$ for the Fermi level positions respective to its optimally doping position. The peaks corresponding to the hyperplasmon of type II and the quasi one-dimensional plasmon are marked by HPII and 1DP, respectively.
The dispersion of the conventional two-dimensional plasmon 2DP at small ${\bf q}$'s outside the electron-hole continuum is not shown.
}
\label{LOSS-2D_DIR-11}
\end{figure*}

\begin{figure}[ht]%
\centering
\includegraphics[width=0.48\textwidth]{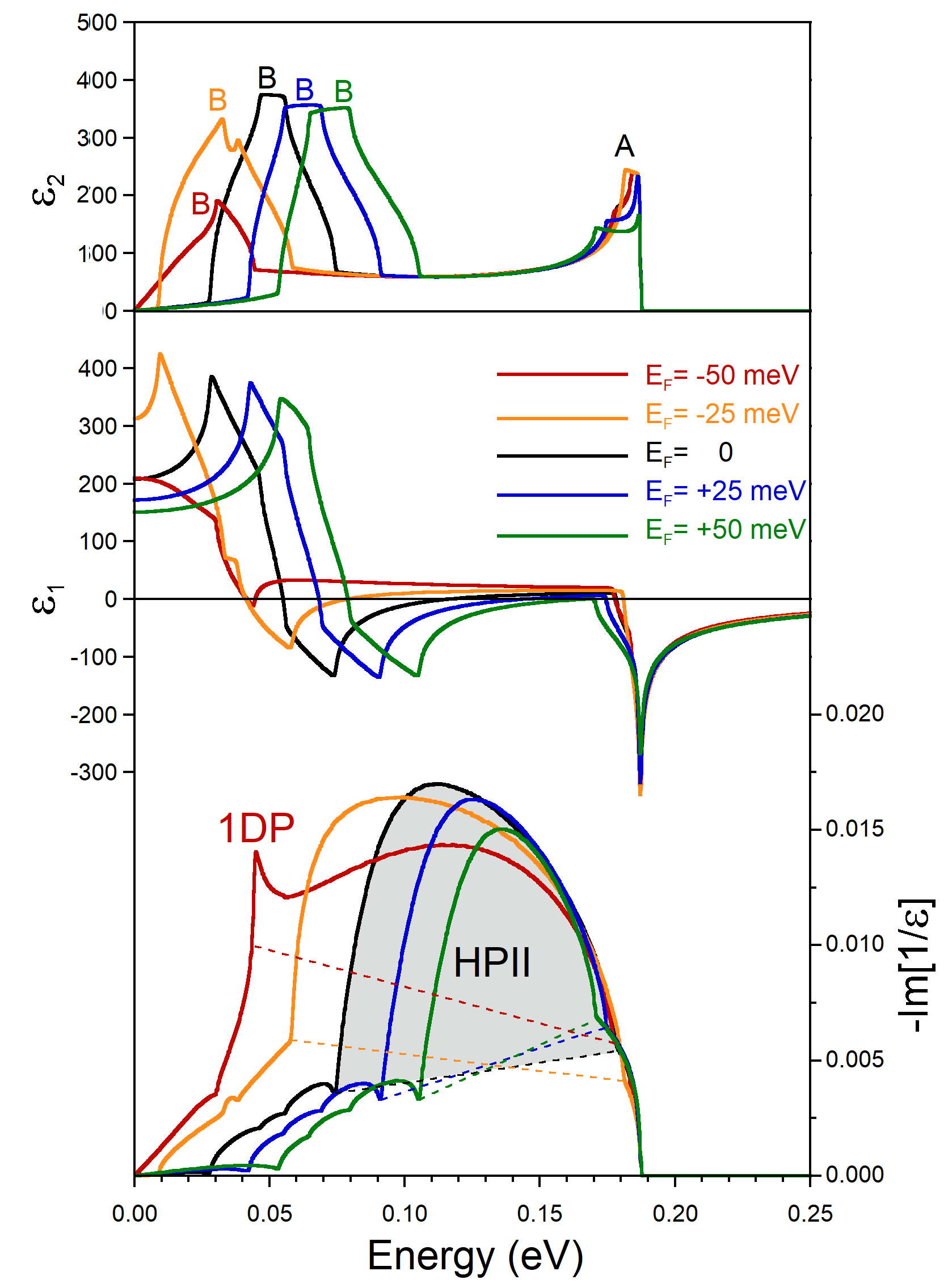}
\caption{ Imaginary (top panel) and real (middle panel) parts of the dielectric function at momentum transfer ${\bf q}=(0.1\pi,0.1\pi)$ for five Fermi-level positions.
 The respective loss functions  are presented in the bottom panel. Spectral weight in the loss function at $E_F=0$ tentatively corresponding to the hyperplasmon HPII is highlighted by grey color.
 Thin dashed lines delimit these regions for all five doping levels. The peak in the loss function at $E_F$=-50 meV corresponding to the quasi-one dimensional plasmon is marked as 1DP.
}
\label{EPS_Q=0250_0250}
\end{figure}

\begin{figure*}[ht]%
\centering
\includegraphics[width=0.95\textwidth]{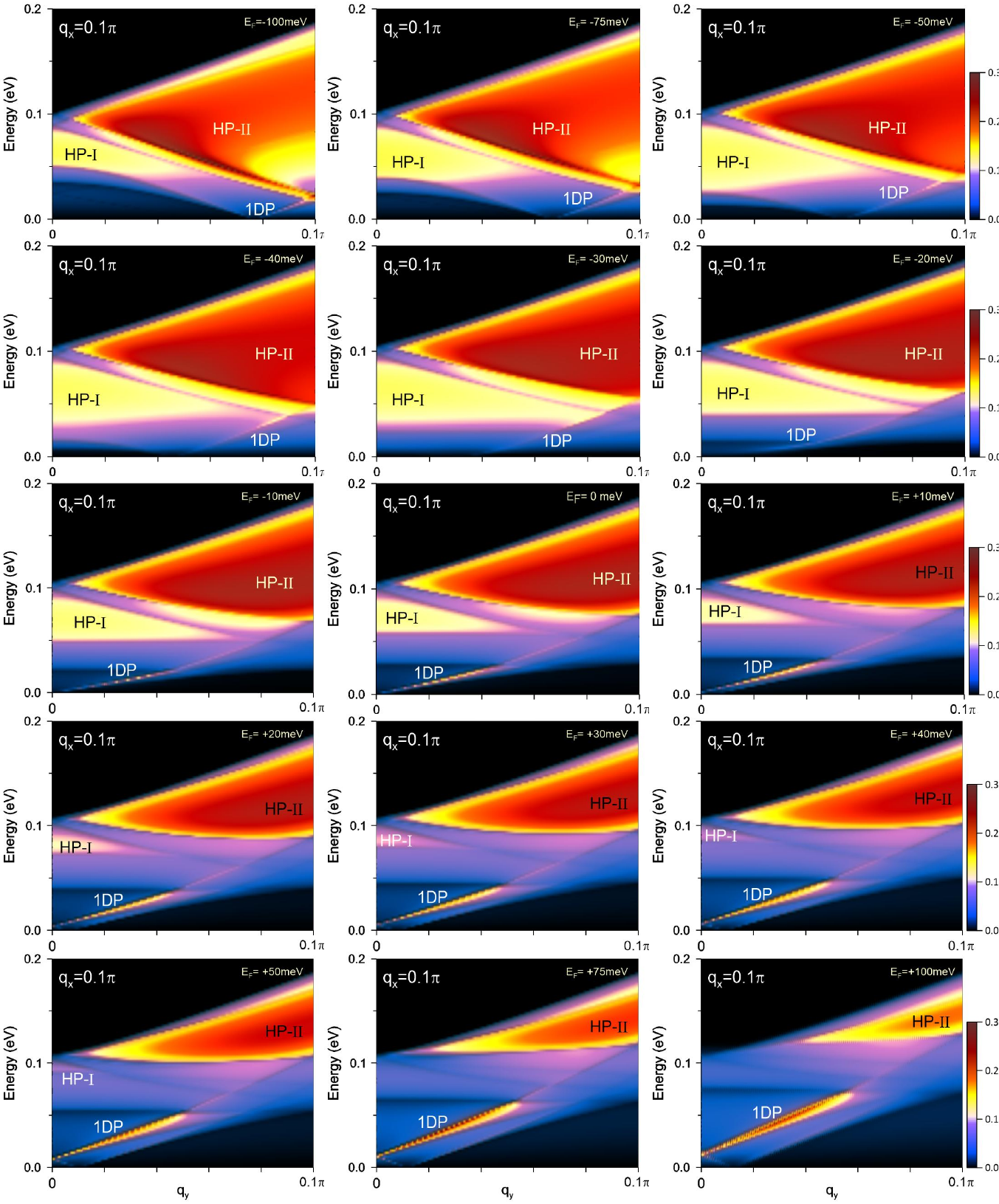}
\caption{ Normalized loss function $L({\bf q},\omega)$=-Im$[1/\epsilon({\bf q},\omega)]/\omega$ at momentum transfers ${\bf q}$ along the $[10]$ direction at fixed $q_x=0.1 \pi$.
Different panels represent $L({\bf q},\omega)$ for the Fermi level positions respective to its optimally doping position. The peaks corresponding to the hyperplasmons of types I and  II,
as well as the quasi one-dimensional plasmon are marked by HPI, HPII, and 1DP, respectively.
}
\label{LOSS-2D_QX=250}
\end{figure*}

 Figure \ref{LOSS-2D_DIR-10} shows that when the Fermi level shifts upward the intensity of the HPI peak quickly reduces. This is accompanied by gradual narrowing of the HPI peak in the loss function. When $E_F$ locates above +50 meV the HPI peak disappears.
 The HPI peak narrowing can be explained by reduction of the energy distance between the A and B peaks in $\epsilon_2$ upon upward shift of $E_F$. For instance, in Fig. \ref{EPS_Q=0250_0000} this distance at $E_F$=+25 meV is significantly smaller than at $E_F$=0, mainly due to the upward shift of the peak B. This shift is caused by strong increase of $v_{F2}$ at small positive energies observed in Fig. \ref{FIG-GV}(a).

 The disappearance of HPI in the loss function is accompanied by emerging of a sharp peak at finite energies corresponding to the 1DP. In the optimally doped case ($E_F$=0),
 along this direction this mode has a zero value from ${\bf q}=(0,0)$ to ${\bf q}(\approx0.4\pi,0)$.\cite{sidrjpcl23} But, as the $E_F$ shifts upward, the 1DP energy becomes
  finite at momenta ${\bf q}$=($q$,0) with $q$'s between zero and some finite value $q_M$ (which depends on the doping) having an arch-like dispersion. For instance, at $E_F$=+100 meV the 1DP disperses from ${\bf q}$=(0,0) to ${\bf q}$=(0.53$\pi$,0), as highlighted by the orange dash line in the bottom-right panel of Fig. \ref{LOSS-2D_DIR-10}. The top energy in the 1DP dispersion increases as well.

In Fig. \ref{LOSS-2D_DIR-11}, the normalized loss functions at different Fermi level positions are reported at momentum transfers ${\bf q}$ along the [11] direction. At $E_F$=0 the loss function is dominated by a prominent HPII peak, dispersing in the energy interval from zero up to about 0.5 eV. This mode have a sound-like dispersion as well. When $E_F$ is shifted downward this peak becomes broader and its intensity increases tiil the $E_F$ reaches a value of  -50 meV. When the Fermi level position is at about -30 meV, a peak corresponding to the 1DP becomes visible in the loss function at small momenta and energies.
 Apparently, it arises gradually on the lower-energy side of the broad HPII peak. With subsequent $E_F$ lowering, the intensity of the 1DP peak increases, its decoupling from the HPII peak becomes more clear, and it expands over the larger momentum region.

As seen in Fig.  \ref{LOSS-2D_DIR-11}, when the Fermi level shifts above zero, the intensity of the HPII peak reduces.  At $E_F$=+100 meV this peak in the loss function becomes very narrow and its spectral weight is extremely small.

In Fig. \ref{EPS_Q=0250_0250} we report the dielectric functions and respective loss functions evaluated at ${\bf q}=(0.1\pi,0.1\pi)$ for five energy positions of the Fermi level. At $E_F$=0, a clear two-peaks structure is observed in the imaginary part of the dielectric function. As discussed above, this results in a small value of $\epsilon_1$ over an extended energy interval between the energy positions of the A and B peaks in $\epsilon_2$.
 The peak B is generated by the intra-band transitions involving the states moving in this direction with the velocities $v_{F1}$ and $v_{F2}$ which  are similar as can be seen in Fig. \ref{FIG-GV}(b). The higher energy peak A is related to the fast carriers moving with the Fermi velocity $v_{F3}$.

When the Fermi level is placed at -25 meV (orange curves) the energy position of the peak A in $\epsilon_2$ is not changed almost. This is explained by a weak dependence of the group velocity of the upper peak in the vicinity of Fermi level  in the DOS reported in Fig. \ref{FIG-GV}(b). On the contrary, a notable shift to the lower energies of the  peak B in $\epsilon_2$ is evident. This is accompanied by a strong increase of the static $\epsilon_1$. The HPII peak in the loss function becomes wider and its spectral weight increases. Moreover, in general, the spectral weight of the loss function increases as well.

At $E_F$=-50 meV (the red curves) the low-energy peak B in $\epsilon_2$ locates at about the same energy as in the $E_F$=-25 meV case. However, its amplitude is notably reduced in comparison to the orange curve. Again, this can be rationalized analyzing the DOS reported in Fig. \ref{FIG-GV}(b). One can see that at the energy of -50 meV the group velocity of the lower peak becomes zero and respective contributions to $\epsilon_2$ are suppressed. At the same time, the velocity of the second peak in DOS at the energy of -50 meV is close to that of the lower peak at -25 meV. At $E_F$=-50 meV, a weak peak B in $\epsilon_2$  produces only a small cusp in $\epsilon_1$ at about 40 meV instead of a shallow zero crossing observed at other four doping levels. This cusp in $\epsilon_1$ results in emerging on the low-energy side of the broad HPII peak of an additional  peak 1DP, corresponding to the quasi one-dimensional plasmon. This appearance of the 1DP is accompanied by reduction of the spectral weight of the HPII peak.

When $E_F$ is placed at +25 meV (blue curves), the low-energy peak B in $\epsilon_2$ shifts upward whereas the peak A stays at fairly the same energy. As a consequence, in the dielectric function the energy region where $|\epsilon_1|\ll\epsilon_2$ reduces. This results in the upward shift of the centre of gravity of the HPII peak in the loss function which is accompanied by reduction of its spectral weight. At $E_F$=+50 meV this tendency maintains.

Analysing Figs. \ref{EPS-TB_FULL_DIR-11}, \ref{LOSS-2D_DIR-11}, and \ref{EPS_Q=0250_0250} one can notice a strong suppression of the loss function on the low-energy side in the optimally-doped case (at $E_F$=0). As a result, a probability of the e-h excitations at the Fermi level in respective momentum regions is extremely low, i.e. some kind of pseudogap appears in a metallic system. Comparing the loss functions at different doping levels one can see how the energy interval with a pseudogap in the loss function in the low-energy side reduces (increases) when $E_F$ shifts to the negative (positive) energy positions. The pseudogap closes completely at $E_F$=-50 meV.

Figure \ref{LOSS-2D_QX=250} demonstrates how the excitation spectra evolve at ${\bf q}=(q_x=0.1\pi,q_y)$ as a function of the $q_y$ coordinate. Here one can see that the HPI and HPII are, indeed, the different modes at all doping levels. Also it is seen
in more details how the 1DP mode peak decouples from the HPII one at large negative  values of the $E_F$.
At small negative $E_F$, the position of the 1DP peak shifts to the right-hand side from its dispersion at $E_F$=0. Consequently, this mode becomes soft at the momenta outside the [10] symmetry directions. On the contrary, at the positive $E_F$ positions the 1DP peak dispersion shifts to the left, in such a way that the energy of this mode becomes positive at $q_y$=0. The resulting full dispersion along the [10] direction can be found  in Fig. \ref{LOSS-2D_DIR-10}.

\section{Discussion}

A problem to find the Pines demon in one-band description of different materials is far from being settled yet though in the two-band systems and in the surface layers (on Tamm levels on the surface) the acoustic plasmons are well-established.
The acoustic plasmon in all the systems corresponds to one of the zeroes of the real part of the dielectric function and is well-defined bosonic quasi-particle.
The discovery\cite{huhun23} of Pines demon in ruthenates Sr$_2$RuO$_4$ where the Fermi-surface has three different sheets (pockets) and partially filled $d$-orbitals triggered investigations of the plasmon spectrum in novel superconducting materials.

It occurs that the dielectric function of the monolayer in the most 2D bismuth family of high-T$_c$ cuprates does not contain the Pines demon. However, when we take into account the full band dispersion it contains very interesting plasmonic spectrum. Besides a standard 2D plasmon with square root q-dependence $\omega_{pl} \sim \sqrt{q}$ of the spectrum we get there two overdamped hyperplasmon modes (contributing to the shape and maximum of the form-factor $S({\bf q},\omega)$) and one mode corresponding to a quasi 1D plasmon which has a linear dispersion in one momentum direction and no dispersion in the other. In contrast to the conventional acoustic plasmon which becomes soft only at the ${\bf q}=0$, the latter has a zero energy over a finite line, i. e. its implication to the low-energy phenomena should be more relevant.

Let us stress that both ruthenates and cuprates are strongly-correlated materials where effective mass substantially exceeds the bare (band) mass ($m^* \sim 4\cdot m$  in  ruthenates\cite{risijpcm95}   and $m^*  \sim 6 \cdot m$ in cuprates) and where Hubbard interaction  is playing a decisive role (Hubbard interaction on $d$-orbitals in cuprates U$_{dd} \sim 6$ eV).\cite{hyscprb89,unfuprb93}
In numerical simulations usually the Mott-Hubbard physics is described in LDA+$U$ or similar schemes which can give reasonable results especially close to a metal-insulator transition.

Throughout both this paper on cuprates and the  publication for ruthenates\cite{huhun23} the authors calculate {\it the uncorrelated dielectric function}. The problem of correctly inserting the $U$ Hubbard term directly in the dielectric function is very nontrivial subject in spite of several recent and early attempts to include Hubbard in the RPA scheme (see e.g., Refs. \onlinecite{gapss78,sharap22}).
Note that already in simple, Hubbard-I approximation\cite{huprsa63} the one-particle Green’s function has a two-pole structure corresponding to Lower and
 Upper Hubbard bands separated by the Hubbard gap of the order of $U$. As a result, the polarization loop $\chi({\bf q},\omega)$ in the RPA scheme
 will contain the contribution from the Upper Hubbard band at large frequencies $\omega \sim U$ (though usually with a small statistical weight of the order of $\frac{\omega}{U}\rightarrow \frac{W}{U}$, $W$ is the bandwidth).

This contribution and the transfer of the statistical weight from small to larger frequencies nevertheless can be important for optical conductivity in the materials
 with moderate $U$-values and not very small $\frac{W}{U}$-ratios. Inclusion of the oxygen orbitals in the consideration complicates the problem switching on additional channels of the statistical weight transfer, connected with the charge transfer gap on top the Hubbard gap (usually the smaller charge transfer gap is inside the large Hubbard gap).\cite{zasaprl85}
An attempt of the better account of Hubbard correlations at least on the level of simple approximations for the dielectric function will be the subject of our next publications.

Finally, considering important subject of superconductivity, note, that, while the Kohn-Luttinger\cite{koluprl65,kachjetpl88,kamipu15} and plasmon mechanisms of superconductivity\cite{ruleprb16,ruleprb17,chprprb19,cawaprb22} compete in low density electron systems, the mechanism of spin-exchange can be dominant in high-T$_c$ cuprates.\cite{karijpcm94}
Nevertheless, the charge sector can be also important in cuprates, at least on the level of the non-trivial corrections to the critical temperature. The low-energy overdamped hyperplasmon modes as well as quasi one-dimensional plasmon mode found in the present paper can contribute in cuprate superconductivity though its statistical weight yet not clear.

Note that an example of the important contribution of the overdamped magnon modes (called paramagnons) to the triplet p-wave superfluidity of He-3 in the framework of the scenario of the almost ferromagnetic Fermi-liquid could encourage us there.\cite{vowo90}
Another example is the contribution of the overdamped spin waves to the singlet d-wave superconductivity of the strongly underdoped high-T$_c$ cuprates close to the antiferromagnetic  ordering in the framework of the slave-fermion approach to the spin liquid state of a slightly doped Mott insulator.\cite{lerpp08,lenarmp06}

\section{Conclusions}

The influence of the doping level on the low-energy dielectric properties in the metallic state of Bi-2212 is studied taking explicitly into account the realistic dispersion of its conducting energy band. Realizing the density response calculations in the framework of the random phase approximation we observed how the properties of two hyperplasmons HPI and HPII having a sound-like dispersion and a quasi one-dimensional plasmon 1DP are changing with the doping level close to the optimal doping.  The hyperplasmons characterized by anysotropic sound-like dispersions  represent overdamped charge collective excitations.  The 1DP  becomes a soft mode over finite momentum interval along the [10] (as well as along [01]) symmetry direction. All these anomalous modes can be important for low-energy physics and contribute strongly to the photoemission spectra and kinetic characteristics of the high T$_c$ cuprates.

Upon increasing the hole doping level (downward shift of the Fermi level) the strength of the hyperplasmons increases and the respective peaks in the loss function become broader. On the contrary, upon upward shift of the Fermi level, these modes becomes weaker. At the $E_F$ position above 50 meV in respect to that at the optimal doping, the hyperplasmon HPI ceases to exist.

As for the 1DP, its spectral weight strongly reduced upon downward shifting of the $E_F$. This is accompanied by its gradual merging the low-energy side of the HPII hyperplasmon peak. Upon reduction of the doping level, the 1DP peak becomes stronger and its energy is finite at all momenta except the zero momentum and some point along the [10] symmetry direction.

\section{Acknowledgments}

V.M.S acknowledges financial support by Grant PID2022-139230NB-I00 funded by MCIN/AEI/10.13039/501100011033.
D.V.E. acknowledges the financial support of DFG (grant numbers 529677299, 449494427) and hospitality of DIPC, San Sebasti\'an.
M.Yu.K. thanks the Program for basic research of the National Research University Higher School of Economics for support.


\end{document}